\documentclass[12pt,axodraw]{article}
\usepackage{epsfig}
\usepackage{amsmath}
\usepackage{amsfonts}
\usepackage{amssymb}
\usepackage{graphicx}
\providecommand{\U}[1]{\protect\rule{.1in}{.1in}}
\begin{document}

\begin{center}
{\Large {\bf The Minimal Supersymmetric Standard Model (MSSM) with $R$-Parity Violation}}\\
M. C. Rodriguez   \\
{\it Grupo de F\'\i sica Te\'orica e Matem\'atica F\'\i sica \\
Departamento de F\'\i sica \\
Universidade Federal Rural do Rio de Janeiro - UFRRJ \\
BR 465 Km 7, 23890-000 \\
Serop\'edica, RJ, Brazil, \\
email: marcoscrodriguez@ufrrj.br \\} 
\end{center}

\date{ April 2022}

\begin{abstract}
In this lectures, we give a review about the Minimal Supersymmetric Standard Model (MSSM) with 
$R$-Parity Violation because it provides an attractive way to generate neutrino masses, 
lepton mixing angles in acconcordance to present neutrino data. 
\end{abstract}

PACS number(s): 12.60. Jv


\section{Introduction}

Although the Standard Model (SM) describes the observed 
properties of charged leptons and quarks it is not the 
ultimate theory. However, the necessity to go beyond it, 
from the experimental point of view, comes at the moment 
only from neutrino data~\cite{bk}.

On September of 2018 from 5-7, happend in Paris, France, the International Conference 
``History of the Neutrino (1930-2018)", where we get the following intersting presentations:
\begin{itemize}
\item Birth of the neutrino, from Pauli to the Reines-Cowan experiment \cite{jarlskog};
\item Prehistory of neutrino oscillation \cite{bilenky};
\item The Nature of the Neutrino (Dirac/Majorana) and Double Beta Decay with or without Neutrinos \cite{petcov};
\item The Mikheyev-Smirnov-Wolfenstein (MSW) Effect \cite{smirnov};
\item Neutrino Mistakes: Wrong tracks and Hints, Hopes and Failures \cite{goodman};
\item Neutrinos and Particle Physics Models \cite{ramond};
\item Neutrino Masses and Mixing: A Little History for a Lot of Fun \cite{gonzales};
\end{itemize}
and many others interestings talks where it was presented in details that neutrinos are massive 
particle, they can oscilate and they can be Dirac particles \cite{Dirac} or 
Majorana fermions \cite{Majorana,{bilenky-majorana}}. A very nice review about the 
status of Neutrinos can be found at \cite{bilenky-status}.

Supersymmetry (SUSY) or symmetry between bosons (particles with integer
spin) and fermions (particles with half-integer spin) has been
introduced in theoretical papers nearly 30 years ago independently by Golfand and Likhtman~\cite{gl},
Volkov and Akulov~\cite{va} and Wess and Zumino~\cite{wz}.
The  supersymmetry algebra was introduced in~\cite{gl} \footnote{They also constructed the first 
four-dimensional field theory with supersymmetry, (massive) quantum electrodynamics of 
spinors and scalars}. There 
and in the Wess-Zumino article~\cite{wz}, the supersymmetry generator $Q$ relates 
bosons with fermions in the usual sense. The Volkov-Akulov article, however, deals only 
with fermions. The supersymmetry generator 
acts in a non-linear way, turning a fermion field into a composite bosonic one made of 
two fermion fields~\cite{va} \footnote{They started the foundations of supergravity}. This illustrates 
that the supersymmetric algebra by itself does not 
require superpartners -- in contrast with what is commonly said or thought now. 
Since that time there appeared thousands of papers.  The reason for this remarkable activity is the unique mathematical
nature of supersymmetric theories, possible solution of various
problems of the SM within its supersymmetric extentions as well as the opening perspective of
unification of all interactions in the framework of a single
theory \cite{dress,Baer:2006rs,Aitchison:2005cf,Rodriguez:2016esw,Rodriguez:2019mwf}. About the 
first version of the Supersymmetric Standard Model see \cite{Fayet:2001xk}.

On this review we will present a short introduction for the Minimal Supersymmetric Standard Model with $R$-Parity violation. First we 
present the model at Sec.(\ref{sec:mssm}), then we present our results when the left handed stau-neutrino get vev, Sec.(\ref{sec:massesl3nonulo}). After it 
we present in a short way the same results when all the left-handed sneutrinos get vev. Our conclusion are found in the last section.

\section{Review of the MSSM with $R$-Parity Interactions.}
\label{sec:mssm}

The Minimal Supersymmetric Standard Model (MSSM), known as MSSM, is a good candidate to be the physics beyond the Standard Model, as presented in several books and review about this subject \cite{dress,Baer:2006rs,Aitchison:2005cf,Rodriguez:2016esw,Rodriguez:2019mwf}. It is the supersymmetric extension of the SM that 
contains a minimal number of states and interactions. The model has the 
gauge symmetry defined as
\begin{equation}
SU(3)_{C} \otimes SU(2)_{L} \otimes U(1)_{Y}
\end{equation} 
extended by the supersymmetry to include the supersymmetric partners of the SM fields which have spins that differ by $+1/2$ as required by the supersymmetric algebra. The MSSM with $R$-Parity conservation has 124 free 
parameters \cite{Baer:2006rs}.

The partner of fermion masses and mixing constitues one of the most important issues in modern physics, the only new fermion that does not mixing is the 
gluinos and the only new bosons that does not mix is the sneutrinos. 

However in the MSSM there are interactions 
that volates Lepton or Baryon Number conservation 
\cite{Moreau:2000hz,Barbier:2004ez}. Some years ago it was proposed a model for the structure of the lepton mixing which 
account for the atmospheric and the solar anomalies. It is based on the simplest one-parameter extension of the MSSM with bi-linear 
$R$-Parity violation 
\cite{valle1,valle2,valle0,valle3,valle4,valle5,valle6,valle7,hall,banks,Romao:1991ex,
Davidson:2000ne,Montero:2001ch}.

We will introduce the following chiral superfields
\begin{eqnarray}
\hat{L}_{iL} &=& \left( 
\begin{array}{c}  
\hat{\nu}_{iL} \\
\hat{l}_{iL}
\end{array}
\right) \sim \left( 1,{\bf 2},-1 \right), \,\ 
\hat{l}^{c}_{iR}\sim \left( 1,{\bf 1},+2 \right), \nonumber \\
, \nonumber \\
\hat{Q}_{iL} &=& \left( 
\begin{array}{c}  
\hat{u}_{iL} \\
\hat{d}_{iL}
\end{array}
\right) \sim \left( 3,{\bf 2}, \frac{1}{3} \right), \,\ 
\hat{u}^{c}_{iR}\sim \left( 3,{\bf 1},- \frac{4}{3} \right), 
\,\ 
\hat{d}^{c}_{iR}\sim \left( 3,{\bf 1}, \frac{2}{3} \right), 
\nonumber \\
\hat{H}_{1} &=& \left( 
\begin{array}{c}  
\hat{H}^{+}_{1} \\
\hat{H}^{0}_{1}
\end{array}
\right) \sim \left( 1,{\bf 2},+1 \right), \,\ 
\hat{H}_{2} = \left( 
\begin{array}{c}  
\hat{H}^{0}_{2} \\
\hat{H}^{-}_{2}
\end{array}
\right) \sim \left( 1,{\bf 2},-1 \right).
\end{eqnarray}
It is possible to rotate the chiral superfields $\hat{H}_{2},\hat{L}_{3L}$ to a new basis 
$\hat{H}^{\prime}_{2},\hat{L}^{\prime}_{3L}$ by a linear transformation given by 
\cite{dress,Baer:2006rs,Davidson:2000ne}
\begin{eqnarray}
\hat{H}^{\prime}_{2}&=&\frac{1}{\mu^{2}+ \mu^{2}_{3}}\left( \mu \hat{H}_{2}- 
\mu_{3}\hat{L}_{3L} \right), \nonumber \\
\hat{L}^{\prime}_{3L}&=&\frac{1}{\mu^{2}+ \mu^{2}_{3}}\left( \mu \hat{H}_{2}+ 
\mu_{3}\hat{L}_{3L} \right).
\end{eqnarray}

The non-zero vaccum expectation value of this model are denoted as
\begin{eqnarray}
\langle H_{1} \rangle &\equiv& \frac{1}{\sqrt{2}}\left[ v_{1}+\sigma^{0}_{1}+ \imath \varphi^{0}_{1} \right], \,\
\langle H_{2} \rangle \equiv \frac{1}{\sqrt{2}}\left[ v_{2}+\sigma^{0}_{2}+ \imath \varphi^{0}_{2} \right], \nonumber \\
\langle \tilde{L}_{i} \rangle &\equiv& \frac{1}{\sqrt{2}}\left[ v^{L}_{i}+ \tilde{\nu}^{R}_{i}+ \imath \tilde{\nu}^{I}_{i} \right].
\end{eqnarray}

The superpotential is given by 
\begin{eqnarray}
W&=& W_{2}+ W_{3},   \\
W_{2}&=&\mu\; \left( \hat{H}_{1}\hat{H}_{2} \right)- \sum_{i=1}^{3}\mu_{i} \left( \hat{L}_{i}\hat{H}_{2} \right),\nonumber \\ 
W_{3}&=& \sum_{i,j,k=1}^{3}\left[\, 
f^{l}_{ij}\left( \hat{H}_{1}\hat{L}_{i}\right) \hat{l}^{c}_{jR}+
f^{d}_{ij}\left( \hat{H}_{1}\hat{Q}_{i}\right) \hat{d}^{c}_{jR}+
f^{u}_{ij}\left( \hat{H}_{2}\hat{Q}_{i}\right) \hat{u}^{c}_{jR}+
\lambda_{ijk} \left( \hat{L}_{i}\hat{L}_{j}\right) \hat{l}^{c}_{kR}
\right. \nonumber \\ &+& \left.
\lambda^{\prime}_{ijk}\left( \hat{L}_{i}\hat{Q}_{j}\right) \hat{d}^{c}_{kR} \right],
\label{suppotMSSM}
\end{eqnarray}
where 
\begin{eqnarray}
\left( \hat{A}\hat{B} \right) \equiv \epsilon_{\alpha \beta} \hat{A}^{\alpha} \hat{B}^{\beta}
\end{eqnarray} 
and we introduce 12 new parameters in addition to the MSSM with 
$R$-Parity conservation model. 

We can add the following soft supersymmetry breaking terms to the MSSM \cite{dress,Baer:2006rs,Aitchison:2005cf} 
\begin{eqnarray}
{\cal L}^{MSSM}_{Soft} &=& {\cal L}^{MSSM}_{SMT} + {\cal L}^{MSSM}_{GMT}+ {\cal L}^{MSSM}_{INT} \,\ ,
\label{The Soft SUSY-Breaking Term prop 2aaa}
\end{eqnarray}
where the scalar mass term ${\cal L}_{SMT}$ is given by the following relation
\begin{eqnarray}
{\cal L}^{MSSM}_{SMT} &=& - \sum_{i,j=1}^{3} \left[\,
\tilde{L}^{\dagger}_{iL} \left( M_{L}^{2}\right)_{ij} \tilde{L}_{jL}+ 
\tilde{l}^{c \dagger}_{iR} \left( M^{2}_{l}\right)_{ij} \tilde{l}^{c}_{jR}+ 
M_{1}^{2} H^{\dagger}_{1}H_{1} + M_{2}^{2} H^{\dagger}_{2}H_{2}
\right] \,\ ,
\label{burro}
\end{eqnarray}
The $3 \times 3$ matrices $M_{L}^{2}$ and $M^{2}_{l}$ are hermitian and $M_{1}^{2}$ and $M_{2}^{2}$ are real. The gaugino 
mass term is written as
\begin{eqnarray}
{\cal L}^{MSSM}_{GMT} &=&- \frac{1}{2}  \left[
\left(\,M_{3}\; \sum_{a=1}^{8} \lambda^{a}_{C} \lambda^{a}_{C}
+ M\; \sum_{i=1}^{3}\; \lambda^{i} \lambda^{i}
+ M^{\prime} \;   \lambda \lambda \,\right)
+ hc \right] \,\ .
\label{The Soft SUSY-Breaking Term prop 3}
\end{eqnarray}
Here, $M_{3},M$ and $M^{\prime}$ are complex. Finally, there is an interaction term ${\cal L}_{INT}$ of the form
\begin{eqnarray}
{\cal L}^{MSSM}_{INT} &=&- B \mu \left( H_{1}H_{2} \right) 
-\sum_{i=1}^{3} \left\{ B_{i} \mu_{i} \left( H_{2}\tilde{L}_{i} \right)
+  \sum_{j,k=1}^{3} \left[ 
\left( H_{1}\tilde{L}_{iL}\right) A^{E}_{ij} \tilde{l}^{c}_{jR}-
\lambda^{\prime}_{ijk}\left( \tilde{L}_{iL}\tilde{Q}_{jL}\right) \tilde{d}^{c}_{kR}
\right] \right\} 
\nonumber \\ &+&hc  \,\ .
\label{burroint}
\end{eqnarray}
The parameters $B \mu$, $B_{i} \mu_{i},A^{E}_{ij}$ and 
$\lambda^{\prime}_{ijk}$ are in general complex \cite{dress,Baer:2006rs}.

\section{Masses}

Here we will present the masses of all particle of this model. The masses of 
gluinos are the same as presented at MSSM with $R$-Parity conservation, due 
this fact we will not present them here, for more detail see .

We first will present the results when $\tilde{L}_{3}\neq 0$, at 
Sec.(\ref{sec:massesl3nonulo}), then the results for the general case, when 
all the left-handed sneutrinos gain vev at Sec.(\ref{sec:masseslinonulo}).

\section{Results with only $\tilde{L}_{3}\neq 0$.}
\label{sec:massesl3nonulo}

\subsection{Bosons Masses}
\label{bosonmassl3}

On the other hand, ${\cal L}_{Higgs}$ give mass to the gauge bosons, throught the following expression: 
\begin{eqnarray}
\left( {\cal D}_{m}H_{1}\right)^{\dagger}\left( {\cal D}_{m}H_{1}\right) + 
\left( {\cal D}_{m}H_{2}\right)^{\dagger}\left( {\cal D}_{m}H_{2}\right) +
\sum_{i=1}^{3} \left( {\cal D}_{m} \tilde{L}_{3}\right)^{\dagger}
\left( {\cal D}_{m} \tilde{L}_{3}\right),
\label{originmassgaugebosons}
\end{eqnarray}
where ${\cal D}_{m}$ is covariant derivates of the SM given by:
\begin{eqnarray}
{\cal D}_{m}H_{1}&\equiv& \partial_{m}H_{1}+ \imath gT^{i}W^{i}_{m}H_{1}+ \imath
g^{\prime} \left( \frac{Y_{H_{1}}}{2} \right)b_{Y^{\prime}}H_{1}, \nonumber \\
{\cal D}_{m}H_{2}&\equiv& \partial_{m}H_{2}+ \imath gT^{i*}W^{i}_{m}H_{2}+ \imath
g^{\prime} \left( \frac{Y_{H_{2}}}{2} \right)b_{Y^{\prime}}H_{2}, \nonumber \\
{\cal D}_{m}\tilde{L}_{i}&\equiv& \partial_{m}\tilde{L}_{i}+ \imath gT^{i*}W^{i}_{m}\tilde{L}_{i}+ \imath
g^{\prime} \left( \frac{Y_{\tilde{L}_{i}}}{2} \right)b_{Y^{\prime}}\tilde{L}_{i}.
\end{eqnarray}
Before calculate expression for the covariant derivatives it is important 
to remember
\begin{eqnarray}
\sum_{i=1}^{3}\sigma^{i}W^{i}_{m}=
\left( 
\begin{array}{cc}  
W^{3}_{m} & \sqrt{2} W^{+}_{m} \\
\sqrt{2} W^{-}_{m} & -W^{3}_{m}
\end{array}
\right), \,\ 
\sum_{i=1}^{3}\sigma^{i*}W^{i}_{m}=
\left( 
\begin{array}{cc}  
W^{3}_{m} & \sqrt{2} W^{-}_{m} \\
\sqrt{2} W^{+}_{m} & -W^{3}_{m}
\end{array}
\right),
\end{eqnarray}
where the charged gauge bosons are defined as 
\begin{equation}
W^{\pm}_{m}= \frac{1}{\sqrt{2}}\left( V^{1}_{m} \mp \imath V^{2}_{m}\right).
\label{wdef}
\end{equation} 

Using those informations we can get
\begin{eqnarray}
{\cal D}_{m} \langle H_{1} \rangle &=& \frac{\imath v_{1}}{2 \sqrt{2}} 
\left( 
\begin{array}{c}  
g \sqrt{2}W^{-}_{m} \\
-gW^{3}_{m}+g^{\prime}W^{\prime}_{m}
\end{array}
\right), \nonumber \\
{\cal D}_{m} \langle H_{2} \rangle &=& \frac{\imath v_{2}}{2 \sqrt{2}}
\left( 
\begin{array}{c}  
-gW^{3}_{m}+g^{\prime}W^{\prime}_{m} \\
-g \sqrt{2}W^{+}_{m}
\end{array}
\right), \nonumber \\
{\cal D}_{m} \langle \tilde{L}_{3} \rangle &=& 
\frac{\imath v^{L}_{3}}{2 \sqrt{2}}
\left( 
\begin{array}{c}  
-gW^{3}_{m}+g^{\prime}W^{\prime}_{m} \\
-g \sqrt{2}W^{+}_{m}
\end{array}
\right),
\end{eqnarray}
from those equations, we can show
\begin{eqnarray}
\left( {\cal D}^{m} \langle H_{1} \rangle \right)^{\dagger}
\left( {\cal D}_{m} \langle H_{1} \rangle \right) &=& 
\frac{v^{2}_{1}}{8} \left[ 2g^{2}W^{+m}W^{-}_{m}+g^{2}W^{3m}W^{3}_{m}-
gg^{\prime}W^{3m}W^{\prime}_{m}-gg^{\prime}W^{\prime m}W^{3}_{m}
\right. \nonumber \\ &+& \left.
(g^{\prime})^{2}W^{\prime m}W^{\prime}_{m}
\right] , \nonumber \\
\left( {\cal D}^{m} \langle H_{2} \rangle \right)^{\dagger}
\left( {\cal D}_{m} \langle H_{2} \rangle \right) &=&
\frac{v^{2}_{2}}{8}
\left[ 2g^{2}W^{-m}W^{+}_{m}+g^{2}W^{3m}W^{3}_{m}-
gg^{\prime}W^{3m}W^{\prime}_{m}-gg^{\prime}W^{\prime m}W^{3}_{m}
\right. \nonumber \\ &+& \left.
(g^{\prime})^{2}W^{\prime m}W^{\prime}_{m}
\right], \nonumber \\
\left( {\cal D}^{m} \langle \tilde{L}_{3} \rangle \right)^{\dagger}
\left( {\cal D}_{m} \langle \tilde{L}_{3} \rangle \right) &=&
\frac{(v^{L}_{3})^{2}}{8}
\left[ 2g^{2}W^{-m}W^{+}_{m}+g^{2}W^{3m}W^{3}_{m}-
gg^{\prime}W^{3m}W^{\prime}_{m}-gg^{\prime}W^{\prime m}W^{3}_{m}
\right. \nonumber \\ &+& \left.
(g^{\prime})^{2}W^{\prime m}W^{\prime}_{m}
\right]. 
\end{eqnarray}

We get the following expression to the charged gauge bosons masses 
\begin{eqnarray}
M^{2}_{W}= \frac{g^{2}}{4} \left( v_{1}^{2}+v_{2}^{2}+(v^{L}_{3})^{2} \right).
\label{wmass}
\end{eqnarray}
We can define the new angle $\beta$ in the following way
\begin{eqnarray}
\tan \beta &=& \frac{v_{2}}{v_{1}},
\label{betadef} 
\end{eqnarray}
and we can define a new angle $\theta$ in the following way
\begin{eqnarray}
v_{1}&=&v \cos \beta \sin \theta , \nonumber \\
v_{2}&=&v \sin \beta \sin \theta , \nonumber \\
v^{L}_{3}&=&v \cos \theta .
\label{newangleatmssmrpv}
\end{eqnarray}
This new angle $\theta$ tends to $( \pi /2)$ rad in the limit of the MSSM 
when $v^{L}_{3}$ go to zero.

The neutral massive gauge boson ($Z^{0}$) get the following mass
\begin{eqnarray}
M^{2}_{Z}&=&
\frac{g^{2}}{4 \cos^{2} \theta_{W}} 
\left( v_{1}^{2}+v_{2}^{2}+(v^{L}_{3})^{2} \right)
= \frac{M^{2}_{W}}{ \cos^{2} \theta_{W}}, 
\label{z-mass}
\end{eqnarray}
where $\theta_{W}$ is the Weinberg angle and it is defined as
\begin{eqnarray}
e=g \sin \theta_{W}=g^{\prime}\cos \theta_{W},
\label{weinbergangledefinition}
\end{eqnarray}
and get a massless foton $A_{m}$. The rotation in this case is
\begin{eqnarray}
\left( \begin{array}{c} A_{m} \\ Z_{m} \end{array} \right)= 
\left(\begin{array}{cc}
\sin \theta_{W} & \cos \theta_{W} \\
\cos \theta_{W} & - \sin \theta_{W}
\end{array} \right)
\left( \begin{array}{c} V^{3}_{m} \\ V_{m} \end{array} \right) \,\ ,
\label{boson5}
\end{eqnarray} 
it is the exact expression we get in the SM. Therefore the neutral 
boson gauge sector is exact the same as in the SM.

\subsection{Charginos Masses}

The supersymmetric partners of the $W^{\pm}$, together with the 
usual charged leptons 
$l_{3L},l^{c}_{3R}$ and the $H^{\pm}$
mix to mass eigenstates called charginos $\chi^{\pm}_{i}$ ($i=1,2,3$) which 
are four--component Dirac fermions. 

From our Lagrangian we get the following terms to get the mass matrix 
for charginos
\begin{eqnarray}
&& M \left( \imath \lambda^{-}\right) \left( \imath \lambda^{+}\right)+ 
\frac{gv_{1}}{\sqrt{2}}\left( \imath \lambda^{+}\right) \tilde{H}^{-}_{1}+ 
\frac{gv_{2}}{\sqrt{2}}\left( \imath \lambda^{-}\right) \tilde{H}^{+}_{2}+
\frac{gv^{L}_{3}}{\sqrt{2}}\left( \imath \lambda^{+}\right) l_{3L}+ 
\mu \tilde{H}^{-}_{1} \tilde{H}^{+}_{2} 
\nonumber \\ &-&
\mu_{3} l_{3L} \tilde{H}^{+}_{2}- 
\frac{f^{\tau}v_{1}}{\sqrt{2}}l_{3L}l^{c}_{3R}+
\frac{f^{\tau}v^{L}_{3}}{\sqrt{2}}\tilde{H}^{-}_{1}l^{c}_{3R}+hc
\label{comantesmassadochargino}
\end{eqnarray}
where
\begin{equation}
\lambda^{\pm} = \frac{1}{\sqrt{2}}\,(\lambda^{1} \mp \imath \lambda^{2}),
\label{wino2comp}
\end{equation}
see definition of $W$ boson given at Eq.(\ref{wdef})  and 
$f^{\tau}\equiv f^{l}_{33}$ and all the fermions fields are two-component
Weyl-van der Waerden fermions~\cite{vdWaerden1,Haber:1994pe,dreiner1}.

In order to get the mass matrix for charginos, we start with the basis
\begin{equation}
\psi^{-} = \left( \imath \lambda^{-} \!,\,  \tilde{H}^{-}_{1} \!,\, l^{-}_{3L} \right)^{T}, \hspace{6mm}
\psi^{+} = \left( \imath \lambda^{+} \!,\, \tilde{H}^{+}_{2} \!,\, 
l^{c}_{3R} \right)^{T}.  
\end{equation}
The mass terms of the lagrangian of the charged 
gaugino--higgsino system can then be written as 
\begin{equation}
{\cal L}_{m} = \frac{1}{2} \, \left( \psi^{+T}\!,\,\psi^{-T} \right)\,
\, Y^{\pm} \,
\left( \begin{array}{c} 
\psi^{+} \\ 
\psi^{-} 
\end{array} \right) + hc
\label{charginosmassmatrixmssmrv}
\end{equation}
where
\begin{equation}
Y^{\pm}= \left( 
\begin{array}{cc} 
0 & X \\ 
X^{T} & 0 
\end{array} 
\right),
\label{y+}
\end{equation}
therefore we can rewrite Eq.(\ref{charginosmassmatrixmssmrv}) in the following way
\begin{equation}
{\cal L}_{m} = \frac{1}{2} \left[ 
\psi^{+T}X\psi^{-}+ \psi^{-T}X^{T} \psi^{+} \right] + hc
\end{equation}
taken into account Eq.(\ref{comantesmassadochargino}), we get the following 
expression
\begin{equation}
X = \left( \begin{array}{ccc} 
M & \frac{gv_{2}}{\sqrt{2}} & 0 \\
\frac{gv_{1}}{\sqrt{2}} & \mu &- \frac{f^{\tau}v^{L}_{3}}{\sqrt{2}} \\
\frac{gv^{L}_{3}}{\sqrt{2}} &- \mu_{3} & \frac{f^{\tau}v_{1}}{\sqrt{2}} 
\end{array} \right).
\label{x}      
\end{equation}

The $3 \times 3$ matrix $X$ of Eq.(\ref{x}) can be put into diagonal form 
in the following way \cite{dress,Baer:2006rs,valle1,valle0}
\begin{equation}
XX^{T}=X^{T}X= \left( \begin{array}{ccc} 
M^{2}+ \frac{g^{2}v^{2}_{2}}{2} & \frac{g}{\sqrt{2}} \left(
Mv_{1}+ \mu v_{2} \right) & \frac{g}{\sqrt{2}} \left(
Mv^{L}_{3}+ \mu_{3} v_{2} \right) \\
\frac{g}{\sqrt{2}} \left(
Mv_{1}+ \mu v_{2} \right) & \frac{1}{2}\left( g^{2}v^{2}_{1}+2 \mu^{2}+
(f^{\tau})^{2}(v^{L}_{3})^{2}
\right) & A  \\
\frac{g}{\sqrt{2}} \left(
Mv^{L}_{3}+ \mu_{3} v_{2} \right) & A & \frac{1}{2}\left( 
(f^{\tau})^{2}v^{2}_{1}+g^{2}(v^{L}_{3})^{2}+2 \mu^{2}_{3}
\right),
\end{array} \right),
\label{prontodiagonalizarcharginos}
\end{equation}
where
\begin{equation}
A=\frac{1}{2}\left( (f^{\tau})^{2}v_{1}v^{L}_{3}+g^{2}v_{1}v^{L}_{3}- 
2 \mu \mu_{3} \right).
\end{equation}

From Eq.(\ref{prontodiagonalizarcharginos}) it is easy to get the following analytical results
\begin{eqnarray}
det(XX^{T})&=&det(X^{T}X)= \frac{(f^{\tau})^{2}}{8}\left[ g^{2}v_{2} 
\left( v^{2}_{1}-(v^{L}_{3})^{2} \right)-2M \left(
\mu v_{1}+ \mu_{3}v^{L}_{3} \right) \right]^{2}, \\
Tr(XX^{T})&=&Tr(X^{T}X)= \frac{1}{2}\left[ 
2(M^{2}+ \mu^{2}+ \mu^{2}_{3})+g^{2}(v^{2}_{1}+v^{2}_{2}+(v^{L}_{3})^{2})+ 
(f^{\tau})^{2}(v^{2}_{1}+(v^{L}_{3})^{2}) \right]. \nonumber \\
\label{valorescharginosantesdadiag}
\end{eqnarray}
Therefore, the smallest eigenvalues of $XX^{T}$ and $X^{T}X$ with tau mass $m_{\tau}$. 

Using $\tan\beta=1$\footnote{See Eq.(\ref{betadef}).} and $M_{Z}=91.187$ GeV, 
$s^{2}_{W}=0.223$\footnote{$\theta_{W}$ is the weak mixing angle and see Eq.(\ref{weinbergangledefinition}).}, 
$\mu_{3}=0.9$ GeV, $v^{L}_{3}=0$ GeV (this value is consistent with that of 
Ref.~\cite{banks,Montero:2001ch}), $\mu=150$ GeV, $M=250$ GeV we obtain from Eq.(\ref{prontodiagonalizarcharginos}) the masses 
$1.777$ (in GeV) for the tau, and 124.3 and 275.7 GeV for the charginos.

\subsection{Neutralinos Masses}
\label{subsec:neutralinos} 

In this review we choose the basis \cite{mssm} 
\begin{equation}
\psi^{0}_{MSSMRPV} = \left( \begin{array}{ccccccc}
\imath \lambda^{\prime} & \imath \lambda^{3} 
& \tilde{H}^{0}_{1} & \tilde{H}^{0}_{2} & 
\nu_{3L}
\end{array} \right)^{\!\rm T}. \hspace{6mm}
\label{neutralinomssm} 
\end{equation}

From our Lagrangian we get the following terms to get the mass matrix 
for neutralinos
\begin{eqnarray}
&& \frac{M}{2} \left( \imath \lambda^{3}\right) \left( \imath \lambda^{3}\right) +
\frac{M^{\prime}}{2} \left( \imath \lambda^{\prime}\right) 
\left( \imath \lambda^{\prime}\right)+ 
\frac{gv_{1}}{\sqrt{2}}\left( \imath \lambda^{3}\right) \tilde{H}^{0}_{1}-
\frac{gv_{2}}{\sqrt{2}}\left( \imath \lambda^{3}\right) \tilde{H}^{0}_{2}+
\frac{gv^{L}_{3}}{\sqrt{2}}\left( \imath \lambda^{3}\right) \nu_{3L}
\nonumber \\ &-&
\frac{g^{\prime}v_{1}}{\sqrt{2}}\left( \imath \lambda^{\prime}\right) 
\tilde{H}^{0}_{1}+
\frac{g^{\prime}v_{2}}{\sqrt{2}}\left( \imath \lambda^{\prime}\right) 
\tilde{H}^{0}_{2}+
\frac{g^{\prime}v^{L}_{3}}{\sqrt{2}}\left( \imath \lambda^{\prime}\right) 
\nu_{3L}- \mu \tilde{H}^{0}_{1} \tilde{H}^{0}_{2} +
\mu_{3} \nu_{3L} \tilde{H}^{0}_{2}+hc
\label{comantesmassadoneutralino}
\end{eqnarray}

The mass terms of the neutral gaugino--higgsino system 
can then be written as 
\begin{equation}
{\cal L}_{m} = \frac{1}{2}\, 
(\psi^{0}_{MSSMRPV})^{\rm T} \, Y^{neutralino}_{MSSMRPV} \, 
\psi^{0}_{MSSMRPV}+ \,\ hc
\label{yodef}
\end{equation}
with 
\begin{equation}
Y^{neutralino}_{MSSMRPV}= \left( \begin{array}{ccccc}
M^{\prime} & 0 &- \frac{g^{\prime}v_{1}}{2}& \frac{g^{\prime}v_{2}}{2} & 
 \frac{g^{\prime}v^{L}_{3}}{2} \\
0 & M & \frac{gv_{1}}{2}&- \frac{gv_{2}}{2} & \frac{gv^{L}_{3}}{2} \\
- \frac{g^{\prime}v_{1}}{2}& \frac{gv_{1}}{2}& 0 &- \mu & 0 \\
\frac{g^{\prime}v_{2}}{2}&- \frac{gv_{2}}{1}&- \mu & 0 & \mu_{3} \\
\frac{g^{\prime}v^{L}_{3}}{2} & \frac{gv^{L}_{3}}{2} & 0 & 
\mu_{3} & 0 
\end{array} \right) \,\ . 
\label{neutralinomssmrv}  
\end{equation}

We can get numerical results for Eq.(\ref{neutralinomssmrv}), using the values presented after Eq.(\ref{valorescharginosantesdadiag}) together 
with $M^{\prime}=-200$ GeV, we obtain a massive neutrino and its mass is 
$m_{\nu_3}=0.051$ eV\footnote{The atmospheric neutrino mass scalae is $m^{2}=3 \cdot 10^{-3} \mbox{eV}^{2}$.}, and four heavy neutralinos with masses 
$269.30,-202.88, -100.55, 84.2$ GeV.

This model break lepton number and therefore necessarily generate non-zero Majorana neutrino masses \cite{dress,Baer:2006rs,valle3,valle4}. At 
tree-level only one of the neutrinos pick up a mass by mixing with neutralinos \cite{valle5,valle6}, leaving the other 
two neutrinos massless \cite{valle7}. The LSP in this case is the neutralino $\tilde{\chi}^{0}_{1}$ and some of two body decay 
are \cite{dress,Baer:2006rs,valle1,valle0}
\begin{eqnarray}
\tilde{\chi}^{0}_{1} \rightarrow \tau^{\pm}W^{\mp}, \,\ \tilde{\chi}^{0}_{1} \rightarrow \nu_{\tau}Z,
\end{eqnarray} 
etc.

\begin{figure}
\centering
\includegraphics[width=0.8\textwidth]{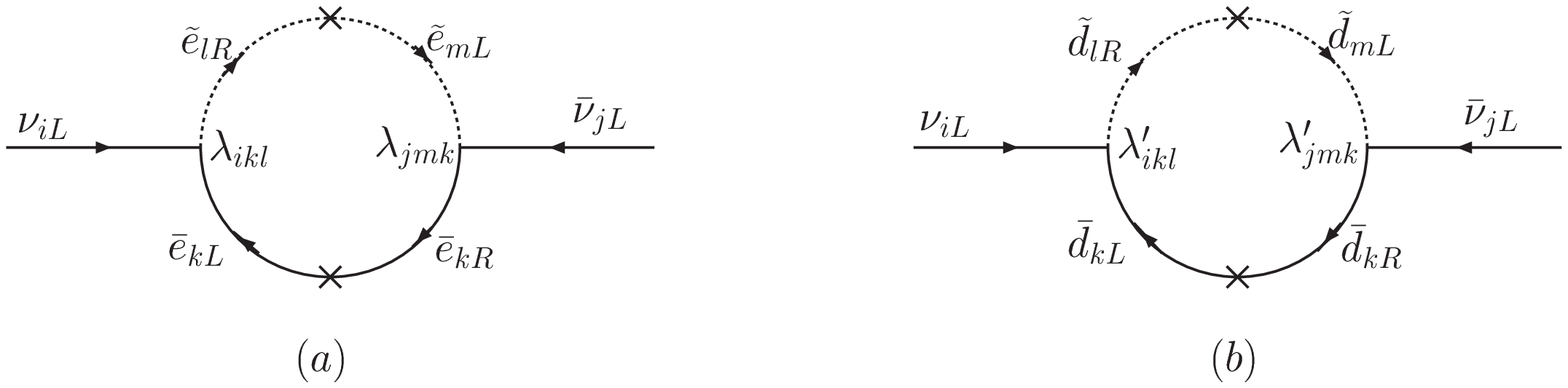}        
\caption{One-loop contribution to neutrino masses, coming from the interaction in our Superpotential defined at Eq.(\ref{suppotMSSM}). This 
figure was taken from \cite{Barbier:2004ez}.} 
\label{f1}
\end{figure}

More realistic neutrino masses require radiative correction and it is shown at Fig.(\ref{f1}) and in this way the solar mass is explained 
and more details about it can be found at \cite{dress,Baer:2006rs,Barbier:2004ez,valle1,hall,banks,Romao:1991ex,Davidson:2000ne,Montero:2001ch}.

We can explain the mixing angle as discussed \cite{valle0}. 

\subsection{Higgs masses}

The Higgs potential of our model has the following form
\begin{equation}
V^{H}_{MSSMRPV}=V_{F}+V_{D}+V_{soft}.
\label{v}
\end{equation}

The first term at Eq.(\ref{v}) is
\begin{equation}
V_{F}=\sum_{l}F_{l}^{\dagger }F_{l},
\end{equation}
where $l=H_{1,2},\tilde{L}_{3L}$ and $\tilde{l}^{c}_{3R}$; the $F$ terms are 
\begin{eqnarray}
F^{\dagger}_{H_{1}}&=&-\frac{\mu}{2}H_{2}- 
\frac{\mu_{3}}{2}\tilde{L}_{3L}, \nonumber \\
F^{\dagger}_{H_{2}}&=&-\frac{\mu}{2}H_{1}-  
\frac{f^{\tau}}{3}\tilde{L}_{3L}\tilde{l}^{c}_{3R}, 
\nonumber \\
F^{\dagger}_{L_{3L}}&=&-\frac{\mu_{3}}{2}H_{1}- 
\frac{f^{\tau}}{3}\tilde{L}_{3L}H_{2}, \nonumber \\
F^{\dagger}_{l_{3R}}&=&- f^{\tau} 
\left( H_{1}\tilde{L}_{3}\right) .
\label{fterms}
\end{eqnarray}
where we have defined $f^{\tau}$ after Eq.(\ref{comantesmassadochargino}) and we get
\begin{eqnarray}
V_{F}&=& \frac{\mu^{2}}{4}\left( 
|H_{1}|^{2}+|H_{2}|^{2} \right)+
\frac{\mu^{2}_{3}}{4}|\left( \tilde{L}_{3L}|^{2}+|H_{2}|^{2} \right) \nonumber \\
&+&
\frac{(f^{\tau})^{2}}{9}\left[ 
\left( |\tilde{L}_{3L}|^{2} + |H_{1}|^{2} \right) |\tilde{l}^{c}_{3R}|^{2}+ 
|H^{\dagger}_{1}\tilde{L}_{3L}|^{2}- |H_{1}|^{2}|\tilde{L}_{3L}|^{2}  \right] 
\nonumber \\ 
&+& 
\frac{\mu_{H}\mu_{3}}{4} \left( 
H^{\dagger}_{2}\tilde{L}_{3L}+ \tilde{L}^{\dagger}_{3L}H_{2} \right)+
\frac{\mu_{H}f^{\tau}}{6} \left( 
H_{1}\tilde{L}^{\dagger}_{3L}\tilde{l}^{c*}_{3R}+
H^{\dagger}_{1}\tilde{L}_{3L}\tilde{l}^{c}_{3R} \right) 
\nonumber \\ &+&
\frac{\mu_{3}f^{\tau}}{6} \left( 
H^{\dagger}_{2}H_{1}\tilde{l}^{c*}_{3R}+
H^{\dagger}_{1}H_{2}\tilde{l}^{c}_{3R} \right), \nonumber \\
\label{vf}
\end{eqnarray}
where $|H_{1}|^{2} \equiv H_{1}^{\dagger}H_{1}$ as usual.

The soft term that contribute to the scalar potential is given by
\begin{eqnarray}  
V_{soft}&=&- {\cal L}_{SMT}-{\cal L}_{Int}.
\label{soft}
\end{eqnarray}
Therefore
\begin{eqnarray}
V_{soft}&=& M^{2}_{H_{1}}|H_{1}|^{2}+ 
M^{2}_{H_{2}}|H_{2}|^{2}+
M^{2}_{L}|\tilde{L}_{3L}|^{2}+ 
M^{2}_{l}|\tilde{l}_{3R}|^{2}+ 
B\mu \left( H_{1}H_{2} \right)+
B_{3}\mu_{3} \left( H_{2}\tilde{L}_{3L} \right) \nonumber \\
&+& 
A^{E}_{33}\left( H_{1}\tilde{L}_{3L} \right) 
\tilde{l}^{c}_{3R}+
hc .
\label{vsoft}
\end{eqnarray}

The third term at Eq.(\ref{v}) is
\begin{equation}
V_{D}=\frac{1}{2}\left[
D^{i}D^{i}+\left( D^{\prime}\right)^{2}
\right],
\end{equation} 
where $i=1,2,3$. There is one $D$-term came from ${\cal L}_{Scalar}$ and from superpotential for each of the four gauge groups
\begin{eqnarray}
SU(2)_{L}: \,\ D^{i}&=&-\frac{g}{2} \left[
H_{1}^{\dagger}\sigma^{i}H_{1}+
H_{2}^{\dagger}\sigma^{i}H_{2}+
\tilde{L}_{3L}^{\dagger}\sigma^{i}\tilde{L}_{3L}
\right], \nonumber \\
U(1)_{Y}: \,\ 
D^{\prime}=&-&\frac{g^{\prime}}{2}\left[
|H_{1}|^{2}-|H_{2}|^{2}- |\tilde{L}_{3L}|^{2}+2 |\tilde{l}^{c}_{3R}|^{2}
\right] .
\label{dterms}
\end{eqnarray}
Using the following relation
\begin{equation}
\sigma^{i}_{ab}\sigma^{i}_{cd}=2 \delta_{ad}\delta_{bc}- \delta_{ab}\delta_{cd},
\end{equation}
we finally get
\begin{eqnarray}
V_{D}&=&\frac{g^{2}}{8}\left[ (|H_{1}|^{2}-|H_{2}|^{2})^{2}+4(
|H^{\dagger}_{2}H_{1}|^{2}+ |H^{\dagger}_{1}\tilde{L}_{3L}|^{2}+
|H^{\dagger}_{2}\tilde{L}_{3L}|^{2})-
2(|H_{1}|^{2}+|H_{2}|^{2}) |\tilde{L}_{3L}|^{2}+
(|\tilde{L}_{3L}|^{2})^{2} \right] 
\nonumber \\ &+& 
\frac{(g^{\prime})^{2}}{8}\left[
|H_{1}|^{2}-|H_{2}|^{2}-|\tilde{L}_{3L}|^{2}+2|\tilde{l}^{c}_{3R}|^{2} \right].
\end{eqnarray}

Using Eqs.(\ref{vf},\ref{vsoft},\ref{dterms}) at Eq.(\ref{v}) we get
\begin{eqnarray}
V^{H}_{MSSMRPV}&=& 
m^{2}_{1h}|H_{1}|^{2}+ 
\left( m^{2}_{2h}+ \mu^{2}_{3} \right)|H_{2}|^{2}+
\left( M^{2}_{L}+ \mu^{2}_{3} \right)|\tilde{L}_{3L}|^{2}+ 
M^{2}_{l}|\tilde{l}^{c}_{3R}|^{2}+ \left[ 
B\mu \left( H_{1}H_{2} \right) \right.
\nonumber \\ &+& \left.
B_{3}\mu_{3} \left( H_{2}\tilde{L}_{3L} \right) - 
A^{E}_{33}\left( H_{1}\tilde{L}_{3L} \right) 
\tilde{l}^{c}_{3R}+ hc \right] \nonumber \\
&+&
\frac{(f^{\tau})^{2}}{9}\left[ 
\left( |H_{1}|^{2}+|\tilde{L}_{3L}|^{2} \right) |\tilde{l}^{c}_{3R}|^{2} -
|H_{1}|^{2}|\tilde{L}_{3L}|^{2}+ 
|H^{\dagger}_{1}\tilde{L}_{3L}|^{2} \right] 
\nonumber \\ 
&+& 
\frac{\mu_{H}\mu_{3}}{4} \left( 
H^{\dagger}_{1}\tilde{L}_{3L}+ \tilde{L}^{\dagger}_{3L}H_{1} \right)+
\frac{\mu_{H}f^{\tau}}{6} \left[ 
(H_{2}\tilde{L}^{\dagger}_{3L})\tilde{l}^{c*}_{3R}+
(H^{\dagger}_{2}\tilde{L}_{3L})\tilde{l}^{c}_{3R} \right] 
\nonumber \\ &+&
\frac{\mu_{3}f^{\tau}}{6} \left[ 
(H^{\dagger}_{2}H_{1})\tilde{l}^{c*}_{3R}+
(H^{\dagger}_{1}H_{2})\tilde{l}^{c}_{3R} \right] + 
\frac{g^{2}}{8}\left[ (|H_{1}|^{2}-|H_{2}|^{2})^{2}+4(
|H^{\dagger}_{2}H_{1}|^{2}+ |H^{\dagger}_{1}\tilde{L}_{3L}|^{2}
\right. \nonumber \\ &+& \left.
|H^{\dagger}_{2}\tilde{L}_{3L}|^{2})-
2(|H_{1}|^{2}+|H_{2}|^{2}) |\tilde{L}_{3L}|^{2}+
(|\tilde{L}_{3L}|^{2})^{2} \right] \nonumber \\
&+& \frac{(g^{\prime})^{2}}{8}\left[
|H_{1}|^{2}-|H_{2}|^{2}-|\tilde{L}_{3L}|^{2}+2|\tilde{l}^{c}_{3R}|^{2}
\right], \nonumber \\
\label{totalpotential}
\end{eqnarray}
where we have defined the following new parameters
\begin{eqnarray}
m^{2}_{1h}&=&M^{2}_{H_{1}} +\frac{\mu^{2}}{4}, \,\
m^{2}_{2h}=M^{2}_{H_{2}} +\frac{\mu^{2}}{4}, \nonumber \\
\end{eqnarray}

From Eq.(\ref{totalpotential}) it is easy to reproduce the minimum of our 
potential scalar and it is given by \cite{dress}
\begin{eqnarray}
V^{min}_{MSSMRPV}&=&\frac{(g^{2}+(g^{\prime})^{2})}{32}
\left( v^{2}_{1}-v^{2}_{2}+(v^{L}_{3})^{2} \right)^{2}+ 
\frac{m^{2}_{1h}}{2}v^{2}_{1}+ \frac{m^{2}_{2h}}{2}v^{2}_{2}-
B\mu v_{1}v_{2}+ B_{3}\mu_{3}v_{2}v^{L}_{3}+ 
\frac{M^{2}_{L}}{2}(v^{L}_{3})^{2} \nonumber \\ &+& 
\frac{\mu^{2}_{3}}{2}(v^{2}_{2}+(v^{L}_{3})^{2})- 
\mu \mu_{3}v_{1}v^{L}_{3},
\end{eqnarray}
and the constraint equataions are
\begin{eqnarray}
m^{2}_{1h}v_{1}-B \mu v_{2}- \mu \mu_{3}v^{L}_{3}+ 
\frac{(g^{2}+(g^{\prime})^{2})}{8}v_{1} \left( v^{2}_{1}-v^{2}_{2}+(v^{L}_{3})^{2} \right)
&=&0, \nonumber \\
\left( m^{2}_{2h}+ \mu^{2}_{3} \right)v_{2}-B \mu v_{1}- B_{3}\mu_{3}v^{L}_{3}+ 
\frac{(g^{2}+(g^{\prime})^{2})}{8}v_{2} \left( v^{2}_{1}-v^{2}_{2}+(v^{L}_{3})^{2} \right)&=&0, \nonumber \\
\left( M^{2}_{L}+ \mu^{2}_{3} \right)v^{L}_{3}- \mu \mu_{3}v_{1}
- B_{3}\mu_{3}v_{2}+ 
\frac{(g^{2}+(g^{\prime})^{2})}{8}v^{L}_{3} \left( v^{2}_{1}-v^{2}_{2}+(v^{L}_{3})^{2} \right)&=&0,
\end{eqnarray}
those equations are in agremment as presented at \cite{dress}.

In the bases
\begin{eqnarray}
&&\left( 
\begin{array}{cccc}  
H^{+}_{1} & H^{+}_{2} & \tilde{l}^{c}_{3R}
\end{array}
\right), \nonumber \\
&&\left( 
\begin{array}{cccc}  
H^{-}_{1} & H^{-}_{2} & \tilde{l}_{3L}
\end{array}
\right),
\end{eqnarray}
the mass matrix is given by
\begin{eqnarray}
M^{2}_{H^{\pm}}&=&
\left( 
\begin{array}{cc}  
M^{2}_{hh} & M^{2}_{\tilde{l}h} \\
M^{2}_{\tilde{l}h} & M^{2}_{\tilde{l}\tilde{l}}
\end{array}
\right),
\end{eqnarray}
where
\begin{eqnarray}
M^{2}_{hh}&=&\left( 
\begin{array}{cc}  
A & B \mu + \frac{g^{2}}{4}v_{1}v_{2} \\
\frac{g^{2}}{4}v_{1}v_{2} & C
\end{array}
\right), \nonumber \\
M^{2}_{\tilde{l}h}&=&\left( 
\begin{array}{cc}  
D & E^{*} \\
E & F
\end{array}
\right), \,\
M^{2}_{\tilde{l}\tilde{l}}=\left( 
\begin{array}{cc}  
G & H \nonumber \\
I & J
\end{array}
\right),
\end{eqnarray}
with
\begin{eqnarray}
A&=&B \mu \frac{v_{2}}{v_{1}}+ \frac{g^{2}}{4}(v^{2}_{2}-(v^{L}_{3})^{2})+ 
\mu \mu_{3}\frac{v^{L}_{3}}{v_{1}}+ \frac{|f^{\tau}|^{2}}{2}(v^{L}_{3})^{2}, \nonumber \\
C&=&B \mu \frac{v_{1}}{v_{2}}+ \frac{g^{2}}{4}(v^{2}_{1}+(v^{L}_{3})^{2})- 
\mu \mu_{3}\frac{v^{L}_{3}}{v_{2}}, \nonumber \\
D&=&\frac{|f^{\tau}|^{2}}{2}v^{2}_{1}- \frac{g^{2}}{4}(v^{2}_{1}-v^{2}_{2})+ 
\mu \mu_{3} \frac{v_{1}}{v^{L}_{3}}-B_{3}\mu_{3}\frac{v_{2}}{v^{L}_{3}}+M^{2}_{L}, 
\nonumber \\
E&=&\frac{f^{\tau}}{\sqrt{2}}\mu_{3}v_{2}\left( A^{E}_{33}+ \mu^{*}\tan \beta \right),
\nonumber \\
F&=&M^{2}_{l}+ \frac{|f^{\tau}|^{2}}{2}(v^{2}_{1}+(v^{L}_{3})^{2})- 
\frac{g^{2}}{4}(v^{2}_{1}-v^{2}_{2}+(v^{L}_{3})^{2}),
\nonumber \\
G&=&- \mu \mu_{3}- \frac{|f^{\tau}|^{2}}{2}v_{1}v^{L}_{3}+ \frac{g^{2}}{4}v_{1}v^{L}_{3}, \nonumber \\ 
H&=&- \frac{1}{\sqrt{2}}\left( f^{\tau}\mu v_{2}-f^{\tau *}A^{E*}_{33}v^{L}_{3} \right), \nonumber \\
I&=&-B_{3}\mu_{3}+ \frac{g^{2}}{4}v_{2}v^{L}_{3}, \nonumber \\ 
J&=&- \frac{f^{\tau *}}{\sqrt{2}}\left( \mu v^{L}_{3}+ \mu_{3}v_{1} \right). 
\end{eqnarray}
those equations are in agremment as presented at \cite{dress}.

We can show the following analytical results
\begin{eqnarray}
det \left( M^{2}_{H^{\pm}} \right)&=&0, \nonumber \\
Tr \left( M^{2}_{H^{\pm}} \right)&=& \frac{1}{v_{1}v_{2}v^{L}_{3}}\left\{
v_{1}v_{2}v^{L}_{3}\left[ 4M^{2}_{l}+2g^{2}v^{2}_{2}-(g^{\prime})^{2}
\left( v^{2}_{1}-v^{2}_{2}+(v^{L}_{3})^{2}\right) \right] +
4B \mu v^{L}_{3}\left( v^{2}_{1}+v^{2}_{2} \right) 
\right. \nonumber \\ &-& \left. 4 \mu_{3} \left[
B_{3}v_{1}\left( v^{2}_{2}+(v^{L}_{3})^{2} \right) +
\mu v_{2} \left( v^{2}_{1}+(v^{L}_{3})^{2} \right)
\right]
\right\}.
\end{eqnarray}

The pseudoscalar squared mass matrix is given
\begin{eqnarray}
M^{2}_{A}=\left( 
\begin{array}{ccc}  
B \mu \frac{v_{2}}{v_{1}}+ \mu \mu_{3}\frac{v^{L}_{3}}{v_{1}} 
& B \mu & - \mu \mu_{3} \\
B \mu & B \mu \frac{v_{1}}{v_{2}}-B_{3}\mu_{3}\frac{v^{L}_{3}}{v_{2}} &-
B_{3}\mu_{3} \\
- \mu \mu_{3} &- B_{3}\mu_{3} & \mu \mu_{3}\frac{v_{1}}{v^{L}_{3}}- 
B_{3}\mu_{3}\frac{v_{2}}{v^{L}_{3}}
\end{array}
\right),
\end{eqnarray}
and we get
\begin{eqnarray}
det \left( M^{2}_{A} \right)&=&0, \nonumber \\
Tr \left( M^{2}_{A} \right)&=& \frac{1}{v_{1}v_{2}v^{L}_{3}}\left\{
4B \mu v^{L}_{3}\left( v^{2}_{1}+v^{2}_{2} \right) - \mu_{3} \left[
B_{3}v_{1}\left( v^{2}_{2}+(v^{L}_{3})^{2} \right) +
\mu v_{2} \left( v^{2}_{1}+(v^{L}_{3})^{2} \right)
\right] \right\}.
\end{eqnarray}
those equations are in agremment as presented at \cite{dress}.

We can also get the mass squared mass matrix for the scalar sector but we will not show it here.

\section{Results with all $\tilde{L}_{i}\neq 0$.}
\label{sec:masseslinonulo}

\subsection{Bosons Masses}

In this case, we continue to using the Eq.(\ref{originmassgaugebosons}) plus 
the following terms 
\begin{eqnarray}
\left( {\cal D}_{m} \tilde{L}_{1}\right)^{\dagger}
\left( {\cal D}_{m} \tilde{L}_{1}\right) +
\left( {\cal D}_{m} \tilde{L}_{2}\right)^{\dagger}
\left( {\cal D}_{m} \tilde{L}_{2}\right),
\label{originmassgaugebosons}
\end{eqnarray}
same the same procedure presented at Sec.(\ref{bosonmassl3}) we get the following expression to the charged gauge bosons masses 
\begin{eqnarray}
M^{2}_{W}= \frac{g^{2}}{4}(v_{1}^{2}+v_{2}^{2}+(v^{L}_{1})^{2}+(v^{L}_{2})^{2}+(v^{L}_{3})^{2}).
\label{wmass}
\end{eqnarray}
then we can rewrite
\begin{eqnarray}
v_{1}&=&v \sin \theta_{1} \sin \theta_{2} \sin \theta_{3}  \sin \beta, \nonumber \\
v_{2}&=&v \sin \theta_{1} \sin \theta_{2} \sin \theta_{3} \cos \beta, \nonumber \\
v^{L}_{3}&=&v \sin \theta_{1} \sin \theta_{2} \cos \theta_{3}, \nonumber \\
v^{L}_{2}&=&v \sin \theta_{1} \cos \theta_{2}, \nonumber \\
v^{L}_{1}&=&v \cos \theta_{1},
\end{eqnarray}
and the neutral massive gauge boson ($Z^{0}$) get the following mass
\begin{eqnarray}
M^{2}_{Z}&=&
\frac{g^{2}}{4 \cos^{2} \theta_{W}} 
(v_{1}^{2}+v_{2}^{2}+(v^{L}_{1})^{2}+(v^{L}_{2})^{2}+(v^{L}_{3})^{2}). 
\label{z-massgeneral}
\end{eqnarray}

\subsection{Charginos Masses}

In this case the $X$ non diagonal mass matrix for charginos, defined at Eq.(\ref{y+}), is given by
\begin{equation}
X = \left( \begin{array}{cc} 
{\cal M}^{MSSM}_{\tilde{\chi}} & {\cal M}_{\tilde{V}L} \\
{\cal M}_{\tilde{H}L}& {\cal M}^{Yukawa}_{lepton}
\end{array} \right).
\label{x}      
\end{equation}
where 
\begin{equation}
{\cal M}^{MSSM}_{\tilde{\chi}} = \left( \begin{array}{cc} 
M & \frac{gv_{1}}{\sqrt{2}} \\
\frac{gv_{2}}{\sqrt{2}} & \mu_{H}
\end{array} \right), \,\       
{\cal M}^{Yukawa}_{lepton} = \frac{v_{1}}{\sqrt{2}} 
\left( \begin{array}{ccc}
f^{l}_{11} & f^{l}_{12} & f^{l}_{13}  \\
f^{l}_{21} & f^{l}_{22} & f^{l}_{23}  \\
f^{l}_{31} & f^{l}_{32} & f^{l}_{33} 
\end{array} \right).
\label{CharginosMSSM;yukawa}      
\end{equation}

The second matrix at Eq.(\ref{x}) is 
\begin{equation}
{\cal M}_{\tilde{V}L} = \left( \begin{array}{ccc}
0 & 0 & 0 \\
A & B & C 
\end{array} \right),
\label{tilde{V}L}      
\end{equation}
where we have defined
\begin{eqnarray}
A&=&- \frac{1}{\sqrt{2}} 
\left(f^{l}_{11}v^{L}_{1}+ f^{l}_{21}v^{L}_{2}+ 
f^{l}_{31}v^{L}_{3} \right), \nonumber \\
B&=&- \frac{1}{\sqrt{2}} 
\left(f^{l}_{12}v^{L}_{1}+ f^{l}_{22}v^{L}_{2}+ 
f^{l}_{32}v^{L}_{3} \right), \nonumber \\
C&=&- \frac{1}{\sqrt{2}} 
\left(f^{l}_{13}v^{L}_{1}+ f^{l}_{23}v^{L}_{2}+ 
f^{l}_{33}v^{L}_{3} \right),
\end{eqnarray}
while the last matrix at Eq.(\ref{x}) has the following expression
\begin{equation}
{\cal M}_{\tilde{H}L} = \left( \begin{array}{cc}
\frac{gv^{L}_{1}}{\sqrt{2}} &- \mu_{1} \\
\frac{gv^{L}_{2}}{\sqrt{2}} &- \mu_{2} \\
\frac{gv^{L}_{3}}{\sqrt{2}} &- \mu_{3} 
\end{array} \right),
\label{tilde{H}L}      
\end{equation}

It is easy to realize that, from Eqs.(\ref{tilde{V}L},\ref{tilde{H}L}), 
chargino sector decouple from the usual lepton sector in the limit 
\begin{eqnarray}
\mu_{1}= \mu_{2}= \mu_{3}=0, \,\ 
v^{L}_{1}= v^{L}_{2}= v^{L}_{3}=0.
\end{eqnarray}

Using the parameters given before Eq.(\ref{valorescharginosantesdadiag}), plus the following parameters 
$f^l_{11}=6.3\cdot10^{-6}$, $f^l_{22}=1.21\cdot10^{-3}$,
$f^l_{33}=2.3\cdot10^{-2}$, $f^l_{12}=f^l_{13}=f^l_{23}=f^l_{21}=f^l_{31}=f^l_{32}=10^{-9}$ we obtain 
from Eq.(\ref{x}) the masses $0.0005,0.105,1.777$ (in GeV) for the usual leptons, and 105.36 and 294.65 GeV for the charginos.

\subsection{Neutralinos Masses}
\label{subsec:neutralinos} 

Using Eq.(\ref{yodef}) we can write in this case the following mass matrix for neutralinos at non-diagnal expression
\begin{equation}
Y^{neutralino}_{MSSMRPV}= \left( \begin{array}{cc}
{\cal M}_{\tilde{\chi}^{0}} & m^{T} \\
m & 0  
\end{array} \right) \,\ . 
\label{neutralinonaodiagonal}  
\end{equation}
where
\begin{equation}
{\cal M}_{\tilde{\chi}^{0}}= \left( \begin{array}{cccc}
M_{1} & 0 &- \frac{g^{\prime}v_{2}}{2}&- \frac{g^{\prime}v_{1}}{2} \\
0 & M_{2} & \frac{gv_{2}}{2}&- \frac{gv_{1}}{2} \\
- \frac{g^{\prime}v_{2}}{2}& \frac{gv_{2}}{2}& 0 &- \mu_{H} \\
\frac{g^{\prime}v_{1}}{2}&- \frac{gv_{1}}{1}&- \mu_{H}& 0  
\end{array} \right) \,\  
m= \left( \begin{array}{cccc}
- \frac{g^{\prime}v^{L}_{1}}{2}& \frac{gv^{L}_{1}}{2}& 0 & \mu_{1}  \\
- \frac{g^{\prime}v^{L}_{2}}{2}& \frac{gv^{L}_{2}}{2}& 0 & \mu_{2} \\
- \frac{g^{\prime}v^{L}_{3}}{2}& \frac{gv^{L}_{3}}{2}& 0 & \mu_{3}   
\end{array} \right) \,\ . 
\label{massaneutrino}  
\end{equation}

Using the same parameters are given after Eq.(\ref{neutralinomssmrv}), but only changing the values 
of $\mu_{3}$ for $\mu_{3}=11.4$,, we obtain a massive neutrino and its mass is 
$m_{\nu_3}=0.051$ eV, and four heavy neutralinos with masses 
$259.05,-208.42, -101.23, 100.65$ GeV. 

\subsection{Higgs masses}

The $F$ terms are 
\begin{eqnarray}
F^{\dagger}_{H_{1}}&=&-\frac{\mu}{2}H_{2}- 
\sum_{i=1}^{3}\frac{\mu_{i}}{2}\tilde{L}_{iL}, \nonumber \\
F^{\dagger}_{H_{2}}&=&-\frac{\mu}{2}H_{1}-  
\sum_{i,j=1}^{3}\frac{f^{l}_{ij}}{3}\tilde{L}_{iL}\tilde{l}^{c}_{jR}, 
\nonumber \\
F^{\dagger}_{L_{iL}}&=&-\frac{\mu_{i}}{2}H_{1}- 
\sum_{j=1}^{3}\frac{f^{l}_{ij}}{3}\tilde{l}^{c}_{jR}H_{2}, \nonumber \\
F^{\dagger}_{l_{iR}}&=&- \sum_{j=1}^{3}\left( f^{l}_{ij}\right)^{2} 
\left( H_{1}\tilde{L}_{jL}\right) .
\label{ftermsgeneral}
\end{eqnarray}

The $D$-term are
\begin{eqnarray}
SU(2)_{L}: \,\ D^{i}&=&-\frac{g}{2} \left[
H_{1}^{\dagger}\sigma^{i}H_{1}+
H_{2}^{\dagger}\sigma^{i}H_{2}+
\sum_{j=1}^{3}\tilde{L}_{jL}^{\dagger}\sigma^{i}\tilde{L}_{jL}
\right], \nonumber \\
U(1)_{Y}: \,\ 
D^{\prime}=&-&\frac{g^{\prime}}{2}\left[
|H_{1}|^{2}-|H_{2}|^{2}- 
\sum_{j=1}^{3}|\tilde{L}_{jL}|^{2}
+2 \sum_{j=1}^{3} |\tilde{l}^{c}_{jR}|^{2}
\right] .
\label{dtermsgeneral}
\end{eqnarray}

The soft term that contribute to the scalar potential is given by
\begin{eqnarray}  
V_{soft}&=&M^{2}_{H_{1}}|H_{1}|^{2}+ 
M^{2}_{H_{2}}|H_{2}|^{2}+
\sum_{i,j=1}^{3} \tilde{L}_{iL} \left( M^{2}_{L} \right)_{ij} \tilde{L}_{jL}+ 
\sum_{i,j=1}^{3} \tilde{l}^{c}_{iR} \left( M^{2}_{l} \right)_{ij} \tilde{l}^{c}_{jR}+ 
B\mu \left( H_{1}H_{2} \right)
\nonumber \\
&+&
\sum_{i=1}^{3}B_{i}\mu_{i} \left( H_{2}\tilde{L}_{3L} \right) + 
\sum_{i,j=1}^{3} 
A^{E}_{ij}\left( H_{1}\tilde{L}_{iL} \right) 
\tilde{l}^{c}_{jR}+
hc .
\label{vsoftgeneral}
\end{eqnarray}

Using Eqs.(\ref{ftermsgeneral},\ref{ftermsgeneral},\ref{vsoftgeneral}), we can reproduce at tree-level the 
scalar mass matrices presented at \cite{valle1,valle0}.

\section{Conclusions}
\label{sec:conclusion}

In this article we have presented the MSSM with $R$-Parity violation. We hope this review can be 
useful to all the people wants to learn about Supersymmetry.

\begin{center}
{\bf Acknowledgments} 
\end{center}
The author would like to thanks to Instituto de F\'\i sica Te\'orica (IFT-Unesp) for their nice 
hospitality during the period I developed this review about SUSY.


\end{document}